\documentclass[%
 aip,
 amsmath,amssymb,
 reprint,%
]{revtex4-1}

\usepackage{graphicx}
\usepackage{dcolumn}
\usepackage{bm}

\usepackage[utf8]{inputenc}
\usepackage[T1]{fontenc}
\usepackage{mathptmx}

\newcommand\Pe{\mbox{Pe}} 
\begin{document}
\preprint{AIP/123-QED}

\title[]{Pair Dispersion in Dilute Suspension of Active Swimmers}
\author{Sergey Belan}
\email{sergb27@yandex.ru}

\author{Mehran Kardar}
\affiliation{ 
Massachusetts Institute of Technology, Department of Physics, Cambridge,
Massachusetts 02139, USA
}


 \begin{abstract}
Ensembles of biological and artificial microswimmers produce long-range velocity fields with strong nonequilibrium fluctuations, which result in dramatic increase diffusivity of embedded particles (tracers).  
While such enhanced diffusivity may point to enhanced mixing of the fluid, a rigorous quantification of the mixing efficiency requires analysis of pair dispersion of tracers, rather than simple one--particle diffusivity.
Here, we calculate analytically the scale-dependent coefficient of relative diffusivity of  passive tracers embedded in a dilute suspension of run-and-tumble microswimmers.
Although each tracer is subject to strong fluctuations resulting in large absolute diffusivity, the small-scale relative dispersion is suppressed due to the correlations in fluid velocity which are relevant when the inter-tracers separation is below the persistence length of the swimmers motion. 
Our results suggest that the reorientation of swimming direction  plays an important role in biological mixing and should be accounted in design of potential active matter devices capable of effective fluid mixing at microscale.   
 \end{abstract}

\maketitle

Suspensions of swimming microorganisms, a prototype of non-equilibrium, exhibit fascinating behaviors distinct from their equilibrium counterparts. 
Even at dilute concentrations, active swimmers can produce large non-Gaussian  fluctuations in fluid velocity with long-range/time correlations, and  not constrained by the fluctuation-dissipation theorem~\cite{Rushkin_2010, Kurihara_2017, Chen_2007, Underhill_2011}.  
The swimmer-induced hydrodynamic fluctuations have been shown to significantly enhance diffusivity of passive tracers placed in active suspensions~\cite{Wu_2000, Leptos_2009, Wilson_2011, Kurtuldu_2011, Jepson_2013, Mino_2013}.  
This observation is relevant to transport of nutrients, and may be relevant to understanding features of bacterial swimming~\cite{Katija_2012}. 
The enhanced tracer diffusion, driven either by motile organisms or by artificial self-propelled particles, can potentially be used for efficient mixing in microfluidic devices~\cite{Kim_2004,Kim_2007}.  

But, does the enhanced diffusivity of tracers in active suspensions actually result in high mixing efficiency?
Diffusive transport and mixing/stirring are often (and in some cases mistakenly, in our opinion) used interchangeably. 
Whereas the former is quantified by the one-particle diffusivity, the latter is associated with the relative dispersion of initially nearby tracers. 
Here we analytically investigate the effect of swimmer-induced hydrodynamic fluctuations on pair dispersion in a dilute active suspension.
Our analysis uncovers the relation between scale-dependent mixing properties and randomization of swimming direction via stochastic tumbling.
If the intertracer separation is sufficiently large, compared to the persistence length of swimmer trajectories, the relative dispersion is determined by the 
absolute diffusivity as the distant tracers move in an uncorrelated manner.
In contrast, on spatial scales below the swimmer persistence length, correlations in the fluid velocity fluctuations can not be neglected leading to weaker relative dispersion.

\begin{figure}
 \includegraphics[scale=0.4]{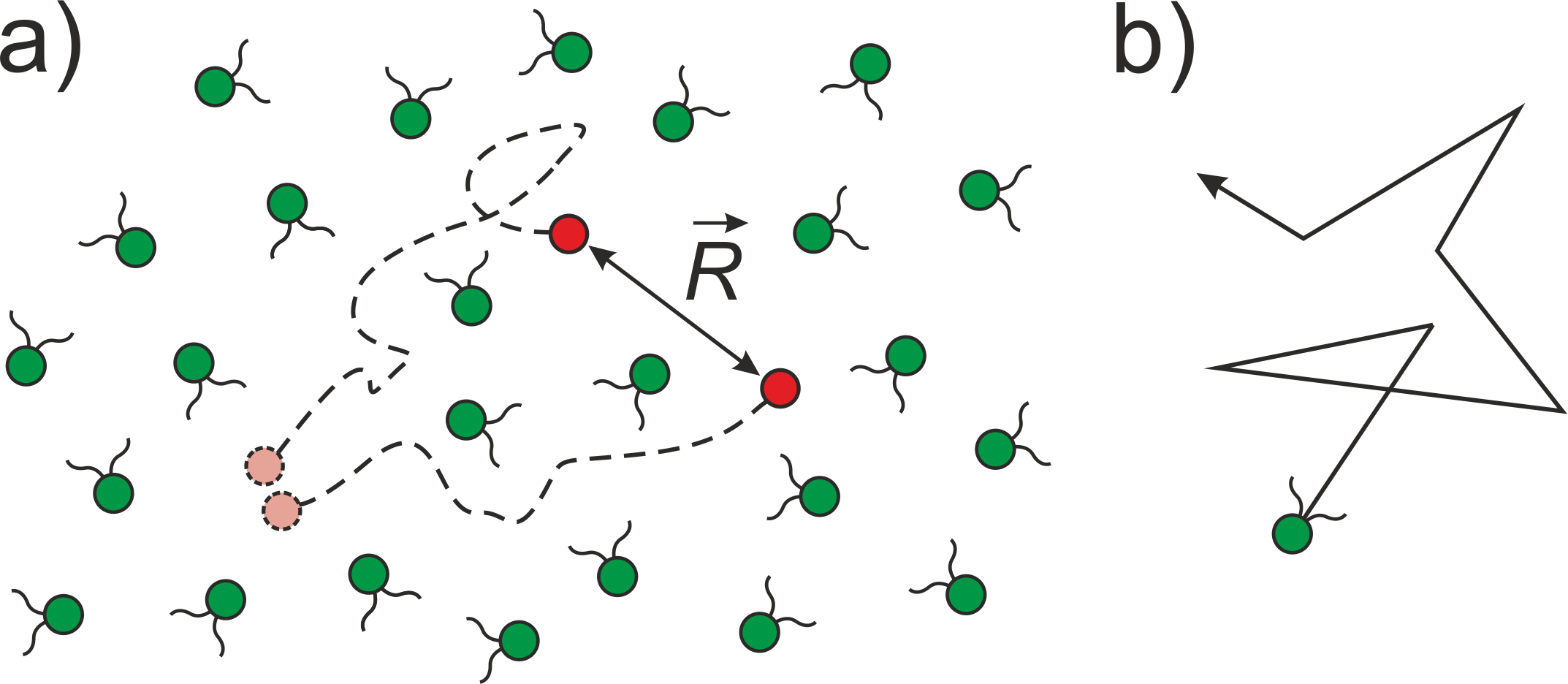}
 \caption{(a): Relative dispersion of passive tracers (dashed line trajectories) in a suspension of active swimmers. (b): Run-and-tumblie motion of an active swimmer.}
  \label{pic:swimmers}
 \end{figure}

Consider two non-Brownian tracer particles moving along  Lagrangian trajectories 
 in the incompressible fluid flow ${\bf v}({\bf r},t)$ produced by an ensemble of active swimmers, see Fig.~\ref{pic:swimmers}a.
Following the seminal work by Richardson~\cite{Richardson_1926}, who established the foundations of two-particle dispersion in hydrodynamic turbulence, we consider the diffusion equation for the probability density  $p({\bf R},t)$ of finding the tracers at separation ${\bf R}$ at time $t$ 
 \begin{equation}
 \label{diffusion_equation}
 \frac{\partial p}{\partial t} = \frac{1}{R^2}\frac{\partial}{\partial R}\left[ R^2D(R)\frac{\partial p}{\partial R}\right],
 \end{equation}
in which the (scale dependent) diffusion coefficient is defined as
 \begin{eqnarray}
 \label{diffusivity_0}
  D(R)=\int\limits_{0}^{\infty} \langle \delta v_{\parallel}({\bf R}, 0)\delta v_{\parallel}({\bf R}, t) \rangle dt,
 \end{eqnarray} 
where  $\delta v_{\parallel}({\bf R},t)=({\bf v}({\bf R},t)-{\bf v}({\bf 0},t))\cdot {\bf R}/R$ represents the Eulerian longitudinal velocity difference along the direction of particle separation, and the angular brackets denote averaging over the statistics of flow fluctuations. 
We assume spatial homogeneity and isotropy of the suspension; 
implying that the relative diffusivity depends only on the absolute value of the 
separation vector.
In addition, Eq.~(\ref{diffusion_equation}) assumes that the probability distribution $p({\bf R},t)$ is spherically symmetric.  
Furthermore, describing the advection of passive tracers by means of a diffusion equation 
is based on the important assumption that the random velocity field is short-range correlated in time.  
This assumption will be justified at the end of the article.

As shown in Refs.~\onlinecite{Batchelor_1953,Kraichnan_1966} 
(see Appendix~\ref{AppA} for the details), the diffusion coefficient defined in Eq.~(\ref{diffusivity_0}) can be written as
\begin{equation}
\label{diffusivity_1}
D(R)=\frac{2}{R^3}\int\limits_{0}^{\infty}\int\limits_0^Rr^2\left(F(0,t)-F(r,t)\right)dtdr,
\end{equation}
in terms of the scalar pair correlation function of the fluid velocity,
\begin{equation}
\label{pair_correlator0}
F(R,t)=\langle {\bf v}({\bf 0},0)\cdot {\bf v}({\bf R},t) \rangle.
\end{equation}
Equation~(\ref{diffusivity_1}) explicitly relates the scale dependencies of relative diffusivity and velocity correlations.  
One may expect that at sufficiently large separations $R$, the term $F(r,t)$ in the integrand of this equation can be neglected as correlations between velocity fluctuations experienced by  distant tracers are almost absent.
Then the coefficient of relative diffusion is simply twice the absolute diffusivity, which can far exceed its thermal value even for  dilute suspensions.
It is clear, however, that the small-scale dispersion must be less pronounced due to the correlations present in the flow.
Indeed, as we will see below,  the relative diffusivity vanishes as the interparticle separation goes to zero. 
What is the characteristic length scale associated with the scale-dependent diffusivity $D(R)$? 
This is the central question of this work.

To calculate $D(R)$, we need to describe the statistics of the flow fluctuations created by the active swimmers.
Since the Reynolds number associated with swimming at microscale is small~\cite{Purcell_1977} ($<10^{-4}$), the fluid motion is governed by the Stokes equation.
For an autonomously moving neutrally buoyant swimmer, a propulsive force is balanced by the resistive drag so that the swimmer exerts no net force on the fluid.
Then the velocity field produced by the swimmer far away from its surface is determined by the leading order dipole term in a multipole expansion.
Noting that the near-field flow (where high order harmonics are relevant) is not universal, 
we will consider dipolar swimmers for simplicity. 
Then, the fluid velocity induced at ${\bf r}$  by a swimmer with orientation ${\bf n}$ placed in the origin is given by (see, e.g., Refs.~\onlinecite{Lauga_2009,Yeomans_2014}) 
\begin{equation}
\label{dipole}
{\bf u}({\bf r},{\bf n})=\frac{\kappa}{8\pi\mu}\left(\frac{3({\bf n}\cdot {\bf r})^2}{r^2}-1\right)\frac{{\bf r}}{r^3},
\end{equation}
where $\kappa$ denotes the strength of the force dipole exerted by the swimmer on the fluid with viscosity $\mu$.
The short-distance cut-off required to regularize singularity at $r\to 0$ is of the order of the physical size $a$ of a swimmer's body.
The strength $\kappa$ is positive for pushers and negative for pullers~\cite{Lauga_2009}. 

\begin{figure}
 \includegraphics[scale=0.38]{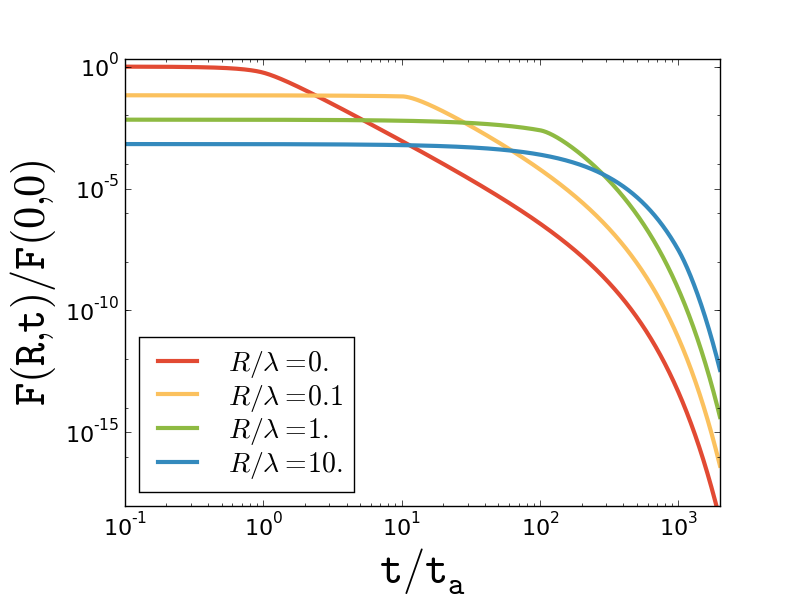}
 \caption{The two-point correlator $F(R,t)$ of the fluid velocity fluctuations as a function of time, for  different spatial separations.  The microscopic time scale $t_a$ is given by $t_a=a/v_0$. The plots correspond to the choice $\alpha t_a=10^{-2}$ (or, equivalently, $a/\lambda=10^{-2}$).}
  \label{pic:F}
 \end{figure}

In the Stokes regime, the total fluid velocity is given by the superposition of flow contributions from all $N\gg 1$ swimmers present in the system, i.e. ${\bf v}({\bf R}, t)=\sum_{k=1}^N {\bf u}({\bf R} - {\bf r}_k(t),{\bf n}_k(t))$, where
 ${\bf r}_k(t)$ and ${\bf n}_k(t)$ indicate, correspondingly, the position and the orientation of the $k$th swimmer, and ${\bf u}$ is given by Eq.~(\ref{dipole}).  
(We consider only a dilute suspension in which interactions between swimmers can
be neglected, and thus assume that they move independently from each other.)
In a statistically stationary state, the swimmers are uniformly distributed in the fluid with a concentration $c$ and the probability distribution $P_0({\bf n})$ of the swimmer orientation vector is isotropic.
Then the two-point correlation function of the fluid velocity defined in Eq.~(\ref{pair_correlator0}) can be written as (see Appendix~\ref{AppC})
\begin{widetext}
\begin{equation}
\label{pair_correlator1}
F(R,t)=c\int {\bf u}(-{\bf r}_1,{\bf n}_1)\cdot{\bf u}({\bf R}-{\bf r}_2,{\bf n}_2)P_0({\bf n}_1)G({\bf r}_2, {\bf n}_2,t|{\bf r}_1,{\bf n}_1,0)~d{\bf r}_1~d{\bf r}_2~d{\bf n}_1~d{\bf n}_2, 
\end{equation}
\end{widetext}
where $G({\bf r}_2, {\bf n}_2,t|{\bf r}_1,{\bf n}_1,0)$ denotes the probability density that after the time $t$ the swimmer will be in  ${\bf r}_2$ with orientation ${\bf n}_2$ having started at ${\bf r}_1$ with the initial swimming direction ${\bf n}_1$.
Equation~(\ref{pair_correlator1}) is valid at the leading order in concentration $c$, accounting for correlations between the positions and orientations of the same swimmer at different moments of time.

To proceed further we need to specify the model for the swimmer's motion. 
Let us assume that each swimmer moves ballistically with  constant speed $v_0$ and undergoes complete reorientation at the rate $\alpha$.
The random reorientations mimic the run-and-tumble behavior of bacteria~\cite{Berg_1993} (see Fig~\ref{pic:swimmers}b).
It is known that real run-and-tumble dynamics is characterized by quick but non-instantaneous tumbles and exhibits correlations between directions of subsequent runs.
Moreover, the runs are not perfectly straight due to rotational diffusion.
However, what is important for succeeding arguments is  that the swimmer's orientation stochastically changes with time, and the fine details of the particular reorientation mechanism are not significant.
As such, we adopt the analytically tractable model of uncorrelated Poisson tumbling to illustrate the main idea.
In this case, the propagator $G$ entering Eq.~(\ref{pair_correlator1}) satisfies the kinetic equation~\cite{Schnitzer_1993} 
\begin{equation}
\label{kinetic}
\partial_t G= -v_0({\bf n}_2\cdot{\bf \nabla}_2) G-\alpha G +\alpha P_0({\bf n}_2)\int G ~d{\bf n}_2',
\end{equation}
supplemented by the initial condition $G({\bf r}_2, {\bf n}_2,0|{\bf r}_1,{\bf n}_1,0)=\delta({\bf r}_2-{\bf r}_1)\delta({\bf n}_2-{\bf n}_1)$.

Equation~(\ref{kinetic}) is equivalent to the Boltzmann equation for the Lorentz model of electron conduction~\cite{Lorentz_1905}.
The exact analytical solution of this equation can be constructed in the Fourier-Laplace space as shown in Refs.~\onlinecite{Hauge_1970,Martens_2012} (see also Appendix~\ref{AppB}).
After lengthy but straightforward calculations, Eq.~(\ref{pair_correlator1}) reduces to  
\begin{equation}
\label{pair_correlator2}
F(R,t)=ce^{-\alpha t}\int {\bf u}(-{\bf r},{\bf n})\cdot{\bf u}({\bf R}-{\bf r}-v_0t{\bf n},{\bf n})P_0({\bf n})~d{\bf r}~d{\bf n}.
\end{equation}
The appealing simplicity of the above relation is due to the vanishing of the contribution  from the last term on the right-hand-side of Eq.~(\ref{kinetic}), see Appendix~\ref{AppC} for the details.

Using Eqs.~(\ref{dipole}) and (\ref{pair_correlator2}), we obtain  the one-point correlation function (see Appendix~\ref{AppD})
\begin{equation}
\label{one_point_correlator}
F(0,t)=\frac{c\kappa^2e^{-\alpha t}}{20\pi\mu^2 a}\left\{\begin{array}{ll}
1-\frac{3}{7}(\frac{v_0t}{a})^2,\ \ t\le \frac{a}{v_0},\\
\\
(\frac{a}{v_0t})^3\left(1-\frac{3}{7}(\frac{a}{v_0t})^2\right),\ \ t>\frac{a}{v_0},
\end{array} \right.
\end{equation}
while the two-point correlation function for $R\gg a$ is found to be (see Appendix~\ref{AppE})
\begin{equation}
\label{two_point_correlator}
F(R,t)=\frac{c\kappa^2e^{-\alpha t}}{30\pi\mu^2R}\left\{\begin{array}{ll}
1,\ \ t\le \frac{R}{v_0},\\
\\
\frac12(\frac{R}{v_0t})^3\left(5-3(\frac{R}{v_0t})^2\right),\ t>\frac{R}{v_0},
\end{array} \right.
\end{equation}
see Fig.~\ref{pic:F} for the illustration. Equations~(\ref{one_point_correlator}) and (\ref{two_point_correlator}) generalize the results of Refs.~\onlinecite{Underhill_2008,Zaid_2011} where the one-point correlator $F(0,t)$ (in the case $\alpha=0$) and the one-time correlator $F(R,0)$ were derived for the dipolar swimmer model.

\begin{figure}
 \includegraphics[scale=0.32]{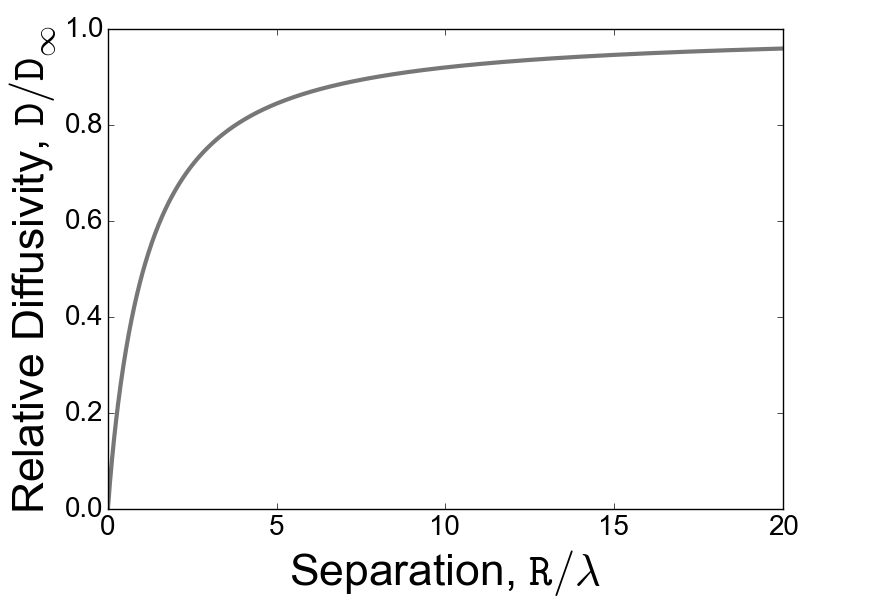}
 \caption{Relative diffusivity as a function of inter-tracers separation.}
  \label{pic:D}
 \end{figure}

Finally, inserting Eqs.~(\ref{one_point_correlator}) and (\ref{two_point_correlator}) into Eq.~(\ref{diffusivity_1}), and performing integration, one finds 
\begin{widetext}
\begin{eqnarray}
\nonumber
D(R)=\frac{c\kappa^2}{24\pi\mu^2v_0}\Big[1+\frac{24(1-e^{-\frac{R}{\lambda}})}{35}\left(\frac{\lambda}{R}\right)^3-\frac{4\lambda}{5R}
-\frac{4e^{-\frac{R}{\lambda}}}{35}\left(\frac{\lambda}{R}\right)^2\left(\frac{3R}{\lambda}+20\right)+\\
\label{diffusivity_2}
+\frac{4\lambda}{R} E_{4}[\frac{R}{\lambda}]+\frac{4}{35}\left(3+56\left(\frac{\lambda}{R}\right)^2\right) E_{5}[\frac{R}{\lambda}]\Big],
\end{eqnarray}
\end{widetext}
where we have introduced the swimmer persistence length $\lambda=\alpha^{-1}v_0$, which is assumed to be large compared to the swimmer's size $a$, and $E_n[x]=\int_1^\infty dt e^{-xt}/t^n$ denotes the  exponential integral.

Equation~(\ref{diffusivity_2}) indicates that the relative diffusivity is a monotonically decreasing function of the interparticle separation,  saturating to $D_{\infty}=\frac{c\kappa^2}{24\pi\mu^2v_0}$ for $R\gg \lambda$, see Fig.~\ref{pic:D}.
Estimating the strength of the force dipole as $\kappa\sim \mu a^2v_0$ (see, e.g., Ref. \onlinecite{Lauga_2009}) we obtain $D_\infty\sim \varphi v_0 a$ where $\varphi=ca^3\ll 1$ is the volume fraction of swimmers.
Obviously, the limit of large separation corresponds to the uncorrelated motion of tracers so that $D_\infty$
is just twice the absolute diffusivity.
Our estimate for $D_{\infty}$ is in agreement with previous studies of the hydrodynamic diffusion of non-Brownian tracers in  dilute three-dimensional suspensions of dipolar swimmers with large persistence length (i.e. $\lambda\gg a$)~\cite{Lin_2011,Pushkin_2013,Kasyap_2014}.

The small-scale asymptotic behavior of the relative diffusivity is linear, $D(R)\approx\frac{D_{\infty}}{\lambda}R$ for $R\ll \lambda$.
Equation~(\ref{diffusion_equation}) then implies a linear growth of the mean interparticle distance with time, $d\langle R(t)\rangle/dt=U$,  with
\begin{equation}
\label{relative_velocity}
U=\frac{c\kappa^2\alpha }{8\pi\mu^2v_0^2},
\end{equation}
playing the role of a mean relative velocity.
Estimated as $U\sim \varphi v_0a/\lambda$, it is much smaller than the (rms) absolute tracer velocity $\sqrt{\langle v^2\rangle}=\sqrt{F(0,0)}\sim \varphi^{1/2}v_0$ as $a/\lambda\ll 1$ and $\varphi\ll 1$ by assumption.  
Despite of the enhanced diffusivity at the single particle level, nearby trajectories diverge relatively slowly due to the underlying velocity correlation. 
Indeed, the typical time it takes for two nearby fluid parcels to reach a separation of 
$R\ll \lambda$  is given by $R/U\sim R\lambda/(\varphi v_0a)$,
much longer than the naive estimate $R^2/D_\infty \sim R^2/(\varphi v_0a)$ based on the assumption that the parcels undergo independent diffusive motions (as they do for $R\gg \lambda$).

As was noted in the beginning of the article, describing dispersion by the diffusion equation is valid
only when the fluid velocity fluctuations are short-range correlated in time.
This is justified since the characteristic time associated with evolution of the separation 
vector accordingly to Eqs.~(\ref{diffusion_equation}) and (\ref{diffusivity_2})  is large compared to the  correlation time of a tracer's relative velocity. 
Using Eqs.~(\ref{one_point_correlator}) and (\ref{two_point_correlator}), the latter can be is estimated  as $\tau_c=\frac{\int_{0}^{\infty} \langle \delta v_{\parallel}({\bf R}, 0)\delta v_{\parallel}({\bf R}, t) \rangle tdt}{\int_{0}^{\infty} \langle \delta v_{\parallel}({\bf R}, 0)\delta v_{\parallel}({\bf R}, t) \rangle dt}\sim \alpha^{-1}$ for $R\ll\lambda$, while $\tau_c\sim a/v_0$  when $R\gg\lambda$.
It is then evident that  $R\lambda/(\varphi v_0a)\gg \alpha^{-1}$ and $R^2/(\varphi v_0a)\gg a/v_0$ and, therefore, the assumption of short correlation times is valid in both limits of small and large tracer separation. 

Noteworthy, setting tumbling rate $\alpha$ to zero we readily find from  Eq.~(\ref{diffusivity_2}) that in the idealized system where swimmers move along infinite straight trajectories the relative diffusivity vanishes.  
This allows us to argue that independently on the particular mechanism underlying the randomization of the swimming direction,  swimmer persistence length $\lambda$ sets the boundary between two regimes of relative dispersion characterized by different mixing efficiencies.
In particular, one may expect that mixing is always suppressed in the systems where swimmer reorientation occurs mainly upon hitting the container walls, since in this case the intertracer separation cannot be larger than the swimmer persistence length which is determined by the system size.

In conclusion, let us discuss some limitations of the analysis presented above, and possible directions for future studies.
First, since we model the swimmer disturbance field as a point force dipole, our results cannot be directly applied to  suspensions of quadrupolar swimmers such as active colloids and certain microorganisms~\cite{Zottl_2016}.
Second, our analysis does not incorporate a detailed model of the near-field hydrodynamic interactions between the swimmer and the tracer particle.
According to a recent experimental study of enhanced diffusion in a suspension of micro-alga~\cite{Kasyap_2014}, tracer entrainment by the near-flow of  swimming microorganisms plays a crucial role in the physical regime when the tracer's size is significantly smaller than that of the swimmers. 
Third, our model ignores  Brownian motion, focusing on the purely convective transport of the tracers.
This restricts applicability of the above results to the limit of large Peclet numbers, $\Pe=D_\infty/D_{th}\gg~1$, where $D_{th}$ is the Brownian diffusivity of tracers.
It may be possible to generalize  calculation of the relative diffusivity beyond the assumptions $\Pe\gg 1$ and $\lambda\gg a$ adopted here,
based on  recent theoretical progress~\cite{Kasyap_2014,Burkholder_2017} in quantifying the absolute diffusivity.
Fourth, here we focused on the very dilute regime neglecting any swimmer-swimmer correlations. 
However, a recent theoretical study~\cite{Stenhammar_2017} indicates that, due to  long-range nature of hydrodynamic interactions between swimmers, 
such correlations can become  significant well below the onset of turbulence, resulting in non-linear scaling of the tracer diffusivity with swimmer concentration.
Thus, further theoretical development is required to extend the present analysis to the case of moderate swimmer densities.

To summarize, we have characterized the spatio-temporal correlations present in a dilute suspension of run-and-tumble microswimmers by calculating the pair correlation function of the fluid velocity fluctuations.
The knowledge of the two-point correlator allows us to derive an analytical expression for the relative diffusivity of passive tracers, thus, revealing those aspects of the mixing process that cannot be captured by the single-particle diffusivity. 
Our results provide insight into the role of swimmer tumbling (and other reorientation mechanisms) in bacteria-induced mixing in natural systems, and should be relevant to design of mixing enhancement systems using active swimmers.

\begin{acknowledgments}
SB gratefully acknowledges support from the James S. McDonnell Foundation via its postdoctoral fellowship
in studying complex systems.
MK acknowledges support from NSF through grant DMR-1708280.
\end{acknowledgments}

\begin{widetext}

\appendix

\section{Relative diffusivity in incompressible random flow}\label{AppA}
  
Here we derive  Eq.~(\ref{diffusivity_1}) for the relative diffusivity of passive tracers.
The probability distribution of the separation vector of two tracers advected by  short-time correlated  incompressible random flow evolves accordingly to the diffusion equation 
\begin{equation}
\label{turbulent_diffusion}
\frac{\partial p}{\partial t}=\frac{\partial}{\partial R_i}\left[D_{ij}(R)\frac{\partial p}{\partial R_j}\right],
\end{equation}
with the diffusivity tensor given by $D_{ij}=\int_0^{+\infty}\langle \delta v_i({\bf R}, t) \delta v_j({\bf R},0)\rangle dt$.
If the flow is statistically homogeneous and isotropic, and the probability distribution $p({\bf R},t)$ is spherically symmetric, then Eq.~(\ref{diffusion_equation}) reduces to (see, e.g., Ref.~\onlinecite{Kraichnan_1966})
\begin{equation}
\frac{\partial p}{\partial t}=\frac{1}{R^2}\frac{\partial}{\partial R}\left[R^2D(R)\frac{\partial p}{\partial R}\right],
\end{equation}
where $D$ is the longitudinal diagonal element of the diffusivity tensor $D_{ij}$ in the coordinate system aligned with ${\bf R}$, which can be expressed as 
\begin{eqnarray}
\label{appendix_diffusivity}
D(R)&=&\int\limits_{0}^{\infty} \langle \delta v_{\parallel}({\bf R}, 0)\delta v_{\parallel}({\bf R}, t) \rangle dt= \int\limits_{0}^{\infty} \langle (v_{\parallel}({\bf R}, 0)-v_{\parallel}({\bf 0}, 0)) (v_{\parallel}({\bf R}, t)-v_{\parallel}({\bf 0}, t)) \rangle dt\nonumber\\
&=&\int\limits_{0}^{\infty} (\langle v_{\parallel}({\bf R}, 0)v_{\parallel}({\bf R}, t)\rangle -\langle v_{\parallel}({\bf R}, 0)v_{\parallel}({\bf 0}, t)\rangle-\langle v_{\parallel}({\bf 0}, 0)v_{\parallel}({\bf R}, t)\rangle+\langle v_{\parallel}({\bf 0}, 0)v_{\parallel}({\bf 0}, t)\rangle) dt \nonumber\\
&=& 2\int\limits_0^\infty (F_{\parallel}(0,t)-F_{\parallel}(R,t))dt,
\end{eqnarray}
in terms of the scalar correlation function  
\begin{eqnarray}
F_{\parallel}(R,t)= \langle  v_{\parallel}({\bf R}, t) v_{\parallel}({\bf 0}, 0) \rangle,
\end{eqnarray}
and relaying on homogeneity and isotropy.

To derive Eq.~(\ref{diffusivity_1}) we note that the general form of the two-point velocity correlator in the homogeneous and isotropic random flow is given by (see Ref.~\onlinecite{Batchelor_1953})
\begin{equation}
\label{correlator_appendix}
F_{ij}(R,t)=\langle v_i({\bf R}, t) v_j({\bf 0},0)\rangle=A(R,t)R_iR_j+\delta_{ij}B(R,t),
\end{equation} 
where $A$ and $B$ are  arbitrary functions of $R$ and $t$.
Let us also introduce the following scalar correlator
\begin{eqnarray}
F_{\perp}(R,t)=\langle v_{\perp}({\bf R}, t) v_{\perp}({\bf 0}, 0) \rangle.
\end{eqnarray} 
Here $v_{\perp}$ is the velocity component along an arbitrary chosen direction orthogonal to the separation vector ${\bf R}$.  
As it follows Eq.~(\ref{correlator_appendix}), the longitudinal and lateral velocity correlation functions, $F_{\parallel}$ are $F_{\perp}$, are related to the functions $A$ and $B$ as
\begin{eqnarray}
F_{\parallel}(R,t)= A(R,t)R^2+B(R,t),\\
F_{\perp}(R,t)=B(R,t),
\end{eqnarray}
and, therefore, Eq.~(\ref{correlator_appendix}) can be rewritten as 
\begin{equation}
\label{correlator_appendix1}
F_{ij}(R,t)=\frac{F_{\parallel}-F_{\perp}}{R^2}R_iR_j+\delta_{ij}F_{\perp}.
\end{equation}
Next, the incompressibility condition, ${\bf \nabla} \cdot{\bf v}=0$, implies that
\begin{eqnarray}
\frac{\partial F_{ij}}{\partial R_i}=\left(\frac{(d-1)(F_{\parallel}-F_{\perp})}{R^2}+\frac{1}{R}\frac{\partial F_{\parallel}}{\partial R}\right)R_j=0,
\end{eqnarray}
where $d$ is the number of spatial dimensions, and consequently
\begin{equation}
\frac{(d-1)(F_{\parallel}-F_{\perp})}{R^2}+\frac{1}{R}\frac{\partial F_{\parallel}}{\partial R}=0.
\end{equation} 
The above equation allows us to express $F_{\perp}$ in terms of $F_{\parallel}$
\begin{equation}
\label{correlator_appendix2}
F_{\perp}=F_{\parallel}+\frac{R}{d-1}\frac{\partial F_{\parallel}}{\partial R}.
\end{equation}
Using Eqs.~(\ref{correlator_appendix1}) and (\ref{correlator_appendix2}), we find  
\begin{equation}
\label{trace_Fij}
F(R,t)=\langle {\bf v}({\bf R}, t)\cdot {\bf v}({\bf 0}, 0) \rangle=F_{ii}(R,t)=F_{\parallel}+(d-1)F_{\perp}=dF_{\parallel}+R\frac{\partial F_{\parallel}}{\partial R},
\end{equation}
and, therefore,
\begin{equation}
\label{correlator_appendix3}
F_{\parallel}(R,t)=\frac{1}{R^{d}}\int\limits_0^RF(r,t)r^{d-1}dr.
\end{equation}
Finally, inserting Eq.~(\ref{correlator_appendix3}) into Eq.~(\ref{appendix_diffusivity}) one obtains
\begin{equation}
\label{appendix_diffusivity1}
D(R)=
\frac{2}{R^d}\int\limits_{0}^{\infty}\int\limits_0^Rr^{d-1}\left(F(0,t)-F(r,t)\right)dtdr.
\end{equation}
This expression gives Eq.~(\ref{diffusivity_1}) for $d=3$.
An equivalent representation of $D(R)$ in terms of the energy-spectrum function (i.e. Fourier transform of $F(r,t)$) can be found in Ref.~\onlinecite{Kraichnan_1966}.

\section{Propagator of the run-and-tumble swimmer}\label{AppB}

Here we construct the solution to Eq.~(\ref{kinetic}) by exploiting the trick proposed in Ref.~\onlinecite{Hauge_1970}.   
It is convenient to use the spherical system of coordinates to parametrize the orientation vector ${\bf n}$.
Then $P_0({\bf n})=\sin\theta/(4\pi)$, where $\theta$ is the polar angle, and Eq.~(\ref{kinetic}) from the main text can be written as 
\begin{equation}
\label{kinetic1}
\partial_t G= -v_0({\bf  n}_2\cdot{\bf \nabla}_2) G-\alpha G +\alpha\frac{\sin\theta_2}{4\pi}\int_{0}^{2\pi}\int_0^\pi G d\varphi_2' d\theta_2'.
\end{equation}
After the Fourier-Laplace transform 
\begin{equation}
\label{G_appendix1}
\tilde G({\bf k}_2,{\bf n}_2, s|{\bf k}_1,{\bf n}_1,0)=\int_{0}^{+\infty}dt e^{-st}\int d{\bf r}_1 d{\bf r}_2 e^{-i{\bf k}_2\cdot{\bf r}_2-i{\bf k}_1\cdot{\bf r}_1}G({\bf r}_2,{\bf n}_2, t|{\bf r}_1,{\bf n}_1,0),
\end{equation}
we obtain
\begin{equation}
\label{G_appendix2}
\tilde G({\bf k}_2,{\bf n}_2, s|{\bf k}_1,{\bf n}_1,0) =\frac{(2\pi)^3\delta(\varphi_2-\varphi_1)\delta(\theta_2-\theta_1)\delta({\bf k}_2+{\bf k}_1)}{\alpha+s+iv_0({\bf n}_2\cdot {\bf k}_2)}+\frac{\alpha \sin\theta_2 \int_{0}^{2\pi}\int_0^\pi \tilde G({\bf k}_2,{\bf n}_2', s|{\bf k}_1,{\bf n}_1,0) d\varphi_2' d\theta_2' }{4\pi(\alpha+s+iv_0({\bf n}_2\cdot {\bf k}_2))}.
\end{equation}
Let us integrate Eq.~(\ref{G_appendix2}) over $d\varphi_2d\theta_2$ assuming that the polar angle $\theta_2$ is measured from the direction determined by the wave vector ${\bf k}_2$
\begin{equation}
\left(1-\frac{\alpha}{4\pi}\int_{0}^{2\pi}\int_0^\pi\frac{\sin\theta_2}{\alpha+s+iv_0({\bf n}_2\cdot{\bf k}_2)}d\varphi_2d\theta_2\right)\int_{0}^{2\pi}\int_0^\pi \tilde G({\bf k}_2,{\bf n}_2', s|{\bf k}_1,{\bf n}_1,0) d\varphi_2' d\theta_2'=\frac{(2\pi)^3\delta({\bf k}_2+{\bf k}_1)}{\alpha+s+iv_0({\bf n}_1\cdot {\bf k}_2)}.
\end{equation}
This allows us to express 
\begin{equation}
\int_{0}^{2\pi}\int_0^\pi \tilde G({\bf k}_2,{\bf n}_2', s|{\bf k}_1,{\bf n}_1,0) d\varphi_2' d\theta_2'=\frac{(2\pi)^3 \delta({\bf k}_2+{\bf k}_1)}{(\alpha+s+iv_0({\bf n}_1\cdot {\bf k}_2))(1-\frac{\alpha}{v_0k_2}\arctan \frac{v_0k_2}{\alpha+s})}.
\end{equation}
Substituting this expression into Eq.~(\ref{G_appendix2}), one obtains 
\begin{equation}
\tilde G({\bf k}_2,{\bf n}_2, s|{\bf k}_1,{\bf n}_1,0)=\frac{(2\pi)^3 \delta({\bf k}_2+{\bf k}_1)}{\alpha+s+iv_0({\bf n}_2\cdot {\bf k}_2)}\left[\delta(\varphi_2-\varphi_1)\delta(\theta_2-\theta_1)+\frac{\alpha\sin\theta_2}{4\pi(\alpha+s+iv_0({\bf n}_1\cdot {\bf k}_2))(1-\frac{\alpha}{v_0k_2}\arctan \frac{v_0k_2}{\alpha+s})} \right],
\end{equation}
or, equivalently, 
\begin{equation}
\label{propagator}
\tilde G({\bf k}_2,{\bf n}_2, s|{\bf k}_1,{\bf n}_1,0)=\frac{(2\pi)^3 \delta({\bf k}_2+{\bf k}_1)}{\alpha+s+iv_0({\bf n}_2\cdot {\bf k}_2)}\left[\delta({\bf n}_2-{\bf n}_1)+\frac{\alpha P_0({\bf n}_2)}{(\alpha+s+iv_0({\bf n}_1\cdot {\bf k}_2))(1-\frac{\alpha}{v_0k_2}\arctan \frac{v_0k_2}{\alpha+s})} \right].
\end{equation}

\section{Velocity correlation function in suspension of swimmers}\label{AppC}

In this section we derive Eqs.~(\ref{pair_correlator1}) and (\ref{pair_correlator2}).
To compute the velocity field of the fluid we need to sum the contributions produced by all swimmers
\begin{equation}
{\bf v}({\bf R}, t)=\sum_{k=1}^N {\bf u}({\bf R} - {\bf r}_k(t),{\bf n}_k(t))=\sum_{k=1}^N\int  \delta({\bf r}-{\bf r}_k(t))\delta({\bf n}-{\bf n}_k(t)) {\bf u}({\bf R}- {\bf r},{\bf n})d{\bf r}~d{\bf n}.
\end{equation} 
Here ${\bf r}_k(t)$ and ${\bf n}_k(t)$ are, respectively, the position and the orientation of the $k$th swimmer, and ${\bf u}$ is given by Eq.~(\ref{dipole}) in the main text. 
Then, the two-point correlation function of the velocity can be written as
\begin{equation}
F(R,t)=\sum_{k,l=1}^N\int  \langle\delta({\bf r}_1-{\bf r}_l(0))\delta({\bf n}_1-{\bf n}_l(0))\delta({\bf r}_2-{\bf r}_k(t))\delta({\bf n}_2-{\bf n}_k(t))\rangle {\bf u}(-{\bf r}_1,{\bf n}_1){\bf u}({\bf R}- {\bf r}_2,{\bf n}_2)d{\bf r}_1~d{\bf r}_2~d{\bf n}_1~d{\bf n}_2.
\end{equation} 

It is easy to check that in the absence of correlations between the different swimmers, all terms with $k\ne l$ go to zero after integration.
Therefore the leading contribution comes from the correlations between position and orientation of the same swimmer at different moments of time.
Taking into account that all swimmers are identical, we then obtain 
\begin{equation}
\label{FRt_appendix1}
F(R,t)=N\int  \langle\delta({\bf r}_1-{\bf r}(0))\delta({\bf n}_1-{\bf n}(0))\delta({\bf r}_2-{\bf r}(t))\delta({\bf n}_2-{\bf n}(t))\rangle {\bf u}(-{\bf r}_1,{\bf n}_1){\bf u}({\bf R}- {\bf r}_2,{\bf n}_2)d{\bf r}_1~d{\bf r}_2~d{\bf n}_1~d{\bf n}_2,
\end{equation}
where we have dropped the label for the swimmer number. 
Next, simple implementation of the Bayes formula yields 
\begin{eqnarray}
\nonumber
&&\langle\delta({\bf r}_1-{\bf r}(0))\delta({\bf n}_1-{\bf n}(0))\delta({\bf r}_2-{\bf r}(t))\delta({\bf n}_2-{\bf n}(t))\rangle\qquad\qquad\\
\label{FRt_appendix2}
\qquad\qquad&&=Pr({\bf r}(0)={\bf r}_1, {\bf n}(0)={\bf n}_1)Pr({\bf r}(t)={\bf r}_2, {\bf n}(t)={\bf n}_2|{\bf r}(0)={\bf r}_1, {\bf n}(0)={\bf n}_1).
\end{eqnarray}
The first term in the right hand side of the last equation represents the joint probability distribution of the
swimmer position and orientation, i.e.
\begin{equation}
\label{FRt_appendix3}
Pr({\bf r}(0)={\bf r}_1, {\bf n}(0)={\bf n}_1)= \frac{1}{V}P_0({\bf n}_1),
\end{equation}
where $P_0$ is the steady-state probability distribution of swimmer orientation  and $V$ (=$N/c$) is the volume of the system.
We used the fact that in the statistically stationary state the probability density of the swimmer's position is uniform at all positions.

The second factor in the right hand side of Eq.~(\ref{FRt_appendix2}) represents the probability density that after the time $t$ the swimmer will be in  ${\bf r}_2$ having the orientation ${\bf n}_2$ provided it starts in ${\bf r}_1$ with orientation ${\bf n}_1$, i.e.
\begin{equation}
\label{FRt_appendix4}
Pr({\bf r}(t)={\bf r}_2, {\bf n}(t)={\bf n}_2|{\bf r}(0)={\bf r}_1, {\bf n}(0)={\bf n}_1)= G({\bf r}_2,{\bf n}_2,t|{\bf r}_1,{\bf n}_1,0).
\end{equation}
From Eq.~(\ref{FRt_appendix1}) together with Eqs.~(\ref{FRt_appendix2}-\ref{FRt_appendix4}) one obtains Eq.~(\ref{pair_correlator1}) from the main text.

Next, using the Fourier-Laplace transform of $G$ and ${\bf u}$
we can rewrite Eq.~(\ref{pair_correlator1}) as
\begin{equation}
\label{FRt_appendix5}
F(R,t)=-\frac{ic}{2\pi}\int_{\gamma}ds e^{st}\int \frac{d{\bf k}_1d{\bf k}_2}{(2\pi)^6}d{\bf n}_1d{\bf n}_2e^{i{\bf k}_2\cdot{\bf R}}\tilde{u}_i({\bf k}_1,{\bf n}_1)\tilde{u}_i({\bf k}_2,{\bf n}_2)P_0({\bf n}_1)\tilde G({\bf k}_2, {\bf n}_2,s|{\bf k}_1,{\bf n}_1,0),
\end{equation}
where $\gamma$ is the contour in the plane of the complex variable $s$, $\tilde{G}$ is given by Eq.~(\ref{propagator}) and
\begin{equation}
\label{dipole_fourier}
\tilde u_i({\bf k},{\bf n})=\int  u_i({\bf r}, {\bf n})e^{-i{\bf k}\cdot{\bf r}}d{\bf r}=\frac{i\kappa}{\mu k^2}\left(\frac{({\bf k}\cdot{\bf n})^2}{k^2}k_i-({\bf k}\cdot{\bf n})n_i\right).
\end{equation}
Taking into account Eq.~(\ref{propagator}), we find 
\begin{eqnarray}
\nonumber
F(R,t)&=&-\frac{ic}{(2\pi)^4}\int d{\bf k} e^{i{\bf k}\cdot{\bf R}}\int\limits_{\gamma}ds e^{st} \int d{\bf n} P_0({\bf n})\frac{\tilde u_i(-{\bf k},{\bf n})\tilde u_i({\bf k},{\bf n})}{\alpha+s+iv_0({\bf n}\cdot {\bf k})}\\
\label{F}
&-&\frac{i\alpha c}{(2\pi)^4}\int d{\bf k}e^{i{\bf k}\cdot{\bf R}}\int\limits_{\gamma}ds e^{st}\int d{\bf n}_1d{\bf n}_2\frac{\tilde u_i(-{\bf k},{\bf n}_1)\tilde u_i({\bf k},{\bf n}_2)P_0({\bf n}_1)P_0({\bf n}_2)}{(\alpha+s+iv_0({\bf n}_1\cdot {\bf k}))(\alpha+s+iv_0({\bf n}_2\cdot {\bf k}))(1-\frac{\alpha}{v_0k}\arctan \frac{v_0k}{\alpha+s})}.
\end{eqnarray}
To perform integration over the swimmer orientations we choose a spherical system of coordinates with the zenith direction parallel to the wave vector ${\bf k}$.
Then the vector ${\bf n}_i$ is parametrised by the angle variables $\theta_i$ and $\varphi_i$.
Let us note that
\begin{eqnarray}
\nonumber
\tilde u_i(-{\bf k},{\bf n}_1)\tilde u_i({\bf k},{\bf n}_2)&=&\frac{\kappa^2}{\nu^2 k^4}\left(\frac{({\bf k}\cdot{\bf n}_1)^2}{k^2}k_i-({\bf k}\cdot{\bf n}_1)n_{1i}\right)\left(\frac{({\bf k}\cdot{\bf n}_2)^2}{k^2}k_i-({\bf k}\cdot{\bf n}_2)n_{2i}\right)\\
&=&\frac{\kappa^2}{\nu^2 k^4}\left(({\bf k}\cdot{\bf n}_1)({\bf k}\cdot{\bf n}_2)({\bf n}_1\cdot{\bf n}_2)-\frac{({\bf k}\cdot{\bf n}_1)^2({\bf k}\cdot{\bf n}_2)^2}{k^2}\right)\nonumber\\
&=&\frac{\kappa^2}{\nu^2 k^2}\cos\theta_1\cos\theta_2 \sin\theta_1\sin\theta_2\cos(\varphi_1-\varphi_2),
\end{eqnarray}
and, therefore, the second term in the rhs of Eq.~(\ref{F}) goes to zero after integration over $\varphi_1$ or $\varphi_2$.
Since this zero contribution comes from the last term on the right hand side of  Eq.~(\ref{kinetic}), and there are no other contributions associated with this term, we can ignore it from the very beginning, thus, passing to the simpler kinetic equation 
\begin{equation}
\label{kinetic2}
\partial_t G= -v_0({\bf  n}_2\cdot{\bf \nabla}_2) G-\alpha G,
\end{equation}
which is exactly solvable and gives  
\begin{equation}
\label{propagator_appendix}
G({\bf r}_2,{\bf n}_2,t|{\bf r}_1,{\bf n}_1,0)=e^{-\alpha t}\delta({\bf r}_2-{\bf r}_1-v_0t{\bf n}_1)\delta({\bf n}_2-{\bf n}_1).
\end{equation}
Substitution of Eq.~(\ref{propagator_appendix}) into Eq.~(\ref{pair_correlator1}) leads to Eq.~(\ref{pair_correlator2}).

\section{The one-point correlation function}\label{AppD}

The one-point correlator $F(0,t)$ can be calculated by inserting the velocity field ${\bf u}$ given by Eq.~(\ref{dipole}) into Eq.~(\ref{pair_correlator2}) and setting $R=0$.
This yields 
\begin{eqnarray}
\nonumber
F(0,t)&=&c\left( \frac{\kappa}{8\pi\mu}\right)^2e^{-\alpha t}\int d{\bf r}d{\bf n} P_0({\bf n})\frac{({\bf r}, {\bf r}+v_0t{\bf n})}{r^3|{\bf r}+v_0t{\bf n}|^3}\left(\frac{3({\bf n}\cdot {\bf r})^2}{r^2}-1\right)\left(\frac{3({\bf n}\cdot (-{\bf r}-v_0t{\bf n}))^2}{|{\bf r}+v_0t{\bf n}|^2}-1\right)\\
\nonumber
&=&c\left( \frac{\kappa }{8\pi\mu}\right)^2e^{-\alpha t}\int d {\bf r} \int_{0}^{2\pi}\frac{d\varphi}{4\pi}\int_{0}^{\pi}d\theta \sin\theta \frac{(r^2+rv_0t\cos \theta)(3\cos^2\theta-1)}{r^3(r^2+v_0^2t^2+2rv_0t\cos\theta)^{3/2}}\left(\frac{3(r^2\cos^2\theta+2rv_0t\cos\theta+v_0^2t^2)}{r^2+v_0^2t^2+2rv_0t\cos\theta}-1\right)\\
\nonumber
&=&\frac{c}{2}\left( \frac{\kappa }{8\pi\mu}\right)^2e^{-\alpha t}\int \frac{d {\bf r}}{r^3} \int_{-1}^{+1}dy \frac{(r^2+rv_0ty)(3y^2-1)}{(r^2+v_0^2t^2+2rv_0ty)^{3/2}}\left(\frac{3(r^2y^2+2rv_0ty+v_0^2t^2)}{r^2+v_0^2t^2+2rv_0ty}-1\right)\\
\nonumber
&=&\frac{c \kappa^2e^{-\alpha t}}{32\pi \mu^2v_0t}\int_a^\infty \frac{d r}{r}\left(\frac{8}{35}\frac{v_0^3t^3}{r^3}(-9+7\frac{r^2}{v_0^2t^2}) h[\frac{r}{v_0t}-1]+\frac{16}{5}\frac{r^2}{v_0^2t^2}(\frac{6}{7}\frac{r^2}{v_0^2t^2}-1) h[1-\frac{r}{v_0t}]\right)\\
\nonumber
&=&\frac{c \kappa^2e^{-\alpha t}}{32\pi \mu^2v_0t}\left(\frac{16}{5}h[1-\frac{a}{v_0t}]\int\limits_{\frac{a}{v_0t}}^{1}  (\frac{6}{7}r^2-1)rdr+\frac{8}{35}\int\limits_{\max(1,\frac{a}{v_0 t})}^{+\infty}\frac{7r^2-9}{r^4}dr\right)\\
&=&\frac{c\kappa^2e^{-\alpha t}}{20\pi\mu^2 a}\left((1-\frac{3}{7}(\frac{v_0t}{a})^2)h[1-\frac{v_0t}{a}] + (\frac{a}{v_0t})^3(1-\frac{3}{7}(\frac{a}{v_0t})^2) h[\frac{v_0t}{a}-1] \right),
\end{eqnarray}
where $h(x)$ is the step function, defined by $h(x)=0$,
$x<0$ and $h(x)=1$, $x\geq 0$.
This result identifies with Eq.~(\ref{one_point_correlator}) in the main text.

\section{The two-point correlation function}\label{AppE}
Here we derive the pair correlator $F(R,t)$ for $R\gg a$.
It is convenient to use the representation given by Eq.~(\ref{FRt_appendix5}).
Taking into account Eqs.~(\ref{dipole_fourier}) and (\ref{propagator_appendix}), we obtain
\begin{eqnarray}
\nonumber
F(R,t)&=&-\frac{ic}{(2\pi)^4}\int d{\bf k} e^{i{\bf k}\cdot{\bf R}}\int\limits_{\gamma}ds e^{st} \int d{\bf n} P_0({\bf n})\frac{\tilde u_i(-{\bf k},{\bf n})\tilde u_i({\bf k},{\bf n})}{\alpha+s+iv_0({\bf n}\cdot {\bf k})}\\
&=& \frac{c\kappa^2 e^{-\alpha t}}{2(2\pi)^3\mu^2}\int\frac{d{\bf k}}{k^2}e^{i{\bf k}\cdot{\bf R}}\int\limits_{0}^{\pi} d\theta\sin^3\theta\cos^2\theta e^{-iv_0tk\cos\theta} = \frac{c\kappa^2 e^{-\alpha t}}{(2\pi)^3\mu^2}\int\frac{d{\bf k}}{k^2}e^{i{\bf k}\cdot{\bf R}}\int_{0}^{1}dy (y^2-y^4)\cos (v_0tky)\nonumber\\
&=&\frac{c\kappa^2e^{-\alpha t}v_0t}{2\pi^2\mu^2R}\int\limits_0^{k_{max}}dk\sin (kR)\frac{(2v_0tk(12-(v_0tk)^2)\cos (v_0tk)+2(5(v_0tk)^2-12)\sin (v_0tk))}{(v_0tk)^6}\nonumber\\
&=&\frac{c\kappa^2 e^{-\alpha t}}{2\pi^2\mu^2R}\int\limits_0^{v_0tk_{max}}dq\sin \left(\frac{qR}{v_0t}\right)\frac{(2q(12-q^2)\cos q+2(5q^2-12)\sin q)}{q^6},
\end{eqnarray}
where $k_{max}\sim a^{-1}$.
At $R\gg a$, we can replace the upper limit of integration by $+\infty$ to get
\begin{eqnarray}
\nonumber
F(R,t)&=&\frac{c\kappa^2 e^{-\alpha t}}{2\pi^2\mu^2R}\int\limits_0^{\infty}dq\sin \left(\frac{qR}{v_0t}\right)\frac{(2q(12-q^2)\cos q+2(5q^2-12)\sin q)}{q^6}\\
&=&\frac{c\kappa^2e^{-\alpha t}}{30\pi\mu^2R} \left( h[1-\frac{v_0t}{R}]+\frac12(\frac{R}{v_0t})^3(5-3(\frac{R}{v_0t})^2)h[\frac{v_0t}{R}-1]\right).
\end{eqnarray}
This result identifies with Eq.~(\ref{two_point_correlator}) in the main text.

\end{widetext}

{}

\end{document}